\documentclass[aps,prb,floats,twocolumn]{revtex4}
\usepackage{epsfig}
\renewcommand{\vec}{\mathbf}
\begin{document}         
\title{The temperature dependence of interlayer
  exchange coupling - spin waves vs. spacer effects}
\author{S.~Schwieger, J. Kienert, and W.~Nolting}
\affiliation{Lehrstuhl Festk{\"o}rpertheorie, Institut f{\"u}r Physik, Humboldt-Universit{\"a}t zu Berlin, 
  Newtonstr. 15, 12489 Berlin}
\begin{abstract}
 There are different mechanisms proposed to be responsible for the
temperature dependence of the interlayer exchange coupling (IEC), namely a
smearing out of the spacer or interface Fermi surface and excitations
and interactions of spin waves. We propose a possibility to separate
both effects by calculating the excitation spectrum of an extended
Heisenberg model and connecting the results with ferromagnetic resonance
(FMR) experiments. To
solve the Heisenberg model we use an approximation that was shown to yield
excellent results. In
this paper the main idea of this procedure is explained and a detailed investigation of the spin wave
contribution to the temperature dependence of FMR
resonance frequencies and fields is carried out.     
\end{abstract}
\maketitle
\section{introduction}
The discovery of the interlayer exchange coupling (IEC) has triggered a renewed interest
in the physics of ultrathin metallic films in the last decades. The term
IEC denotes the coupling of magnetic layers separated by non magnetic spacer
layers that occurs in quite a number of metallic and semiconducting
systems.\cite{Bru95,MVU97} The general mechanism of the coupling and a
lot of additional features are well understood today. The coupling is
caused by spin dependent reflection of spacer electrons at the
magnet/non-magnet interfaces leading to a spin dependent interference
and therewith to a spin dependent renormalization of the density of
states (DOS) and the free energy within the spacer.\cite{Bru95} 
The modification of the
spacer DOS and energy depends
sensitively on the relative orientation of the magnetic layers favoring
a special configuration. This results in the interlayer coupling.
The coupling may be ferro- or antiferromagnetic and oscillates with respect
to the spacer thickness. The periods of the oscillation depend on the
spacer Fermi surface, while the amplitudes as well as the phases may be
additionally influenced by interface roughness, interdiffusion,
disorder, and related effects.\\
However, up to now it is not clear which kind of process dominates the
 temperature dependence of the coupling. There are several candidates
 which are evaluated and compared in Ref.~\onlinecite{ScN04}:
\begin{itemize}
\item[(i)] One source of temperature dependence is connected with the
  described coupling mechanism and is caused by the smearing out of the
  Fermi-surface of the spacer.\cite{BrC91,EMM91} Furthermore temperature
  influences  the reflection coefficients at the magnet/non-magnet interfaces.\cite{AMV96} Both processes
  influence the energy of the spacer layer and will be called
  ``spacer effect'' in the following. 
\item[(ii)] Another source of temperature dependence are magnetic
  excitations, altering the properties of the magnetic
  layers\cite{AMT95} (``magnetic layer effect''). Thermal
  magnetic disorder may drastically reduce the energy difference between
  parallel and antiparallel alignment of the magnetic layers and
  therewith the interlayer coupling.  
\end{itemize}
To find the dominating mechanism elaborate ferromagnetic resonance (FMR)
measurements were performed at a prototype trilayer,\cite{LiB03,LiB03a}
namely a ${\rm Ni_7/Cu_x/Co_2/Cu(001)}$ system where the
subscripts denote the number of atomic layers of each film. The data are analyzed using a classical
continuum model, where the classical equation of motion for the
magnetization \cite{Kit47} or an expansion of the free
energy\cite{SmB55} is considered. This analysis gives the temperature
dependence of an effective interlayer coupling $\tilde{J}_I(T)$ which is
discussed in detail in Ref.~\onlinecite{LRK02} and reconsidered in
Ref.~\onlinecite{ScN04}. This quantity is proportional to the energy
difference between the state where the magnetization of
both magnetic films are parallel to each other and the state where the magnetization of the lower magnetic film are aligned antiparallel to the
magnetization of the upper one:
\begin{equation}
\tilde{J}_I(T)\sim F_{\uparrow\uparrow}(T)-F_{\uparrow\downarrow}(T).
\label{tilde_Ji}
\end{equation}
Note that the magnetic moments are supposed to be parallel within the
magnetic films.
 The most important feature of $\tilde{J}_I(T)$ found experimentally is
 its effective
$T^{3/2}$-like decrease with temperature. Naturally both temperature dependencies, i.e. the
temperature dependence due to the Fermi surface softening and those due
to magnetic excitations, are included in $\tilde{J}_I(T)$.
Thus, since both mechanisms (i) and (ii) contribute,
the dominating mechanism cannot be deduced directly from $\tilde{J}_I(T)$.
\cite{LRK02,ScN04}\\
On the other hand a procedure
sensitive to only one of the mechanisms (i) or (ii) would be highly
desirable. It would allow to separate the mechanisms and to deduce the source of temperature
dependence of the IEC directly from the measurements. It is the intention
of this paper to show that such a procedure exists. The new approach is
based on an alternative analysis of FMR measurements using the microscopic
Heisenberg model.\\ To illustrate the idea let us consider a modeling
of the considered trilayer systems:
\begin{eqnarray}
\label{model-rough}
H&=&-
\sum_{\langle ij \rangle \alpha}J_{\alpha}\vec{S}_{i\alpha}\vec{S}_{j\alpha}
-\sum_{\langle i\alpha\,j\beta\rangle}^{\alpha\neq\beta} J_I\vec{S}_{i\alpha}\vec{S}_{j\beta}\nonumber\\
&&-\sum_{i\alpha}g_{J\alpha}\mu_B\vec{B}_0\vec{S}_{i\alpha}
-\sum_{i\alpha} K_{2\alpha}S_{i\alpha z}^2\\
&&+\sum_{ij\alpha}g_{0\alpha\beta}\left(\frac{1}{r_{ij}^3}\vec{S}_{i\alpha}\vec{S}_{j\beta}-\frac{3}{r_{ij}^5}(\vec{S}_{i\alpha}\vec{r}_{ij})(\vec{S}_{j\beta}\vec{r}_{ij})\right),\nonumber
\end{eqnarray} 
where $\alpha$ runs over the two different layers while $i$ and $j$
denote a lattice site within a layer. For convenience the sums in the
first lines are taken over nearest neighbors only. Thus the {\it intra}layer
coupling parameters $J_\alpha$ as well as the {\it inter}layer coupling
parameter $J_I$ are effective parameters containing also the
contributions of non-nearest-neighbor interactions.
The model describes two magnetic layers coupled by the
interlayer coupling $J_I$, which may be positive or negative. The intralayer coupling constants
$J_{\alpha}$ are chosen to be positive to describe ferromagnetic layers. The
model further accounts for anisotropies due to spin orbit 
($K_{2\alpha}$) and dipolar ($g_{0\alpha}$) coupling as well as for an external field $g_{J\alpha}\mu_B\vec{B}_0$ applied in arbitrary
direction. The parameter $g_{J\alpha}$ denotes the Land${\rm \acute{e}}$
factor of the layer $\alpha$. The
anisotropy due to spin-orbit coupling $K_{2\alpha}$, which includes
magnetocrystalline as well as magnetoelastic contributions,\cite{Bab96} will often be termed
"lattice anisotropy" in the following.
The film plane is the $xy$-plane and tetragonal symmetry is
assumed. Such a model is often used to describe magnetic properties of
transition metal films,\cite{JeB98} especially in the presence of
anisotropies.
Note that the spacer of the trilayer system is not considered
explicitely but its effects on the magnetic layers are described by the
model parameter $J_I$. The latter is proportional to the energy
difference between the states of parallel and antiparallel
aligned magnetizations in both layers,
$\vec{M}_a\,\uparrow\uparrow\,\vec{M}_b$ and $\vec{M}_a\,\uparrow\downarrow\,\vec{M}_b$, if saturated magnetism is assumed. This applies for $T=0$ or at finite
temperatures for cases
where the spin wave excitations of the model (\ref{model-rough}) are
artificially frozen. For the latter case a temperature dependence of
$J_I$ can be caused exclusively by excitations within the spacer. Hence, $J_I(T)$ is proportional
to the difference between the energy contributions {\it of the spacer} in the parallel and antiparallel state:
\begin{equation}
{J}_I(T)\sim F_{\uparrow\uparrow}^{sp}(T)-F_{\uparrow\downarrow}^{sp}(T).
\label{Ji}
\end{equation}
\\
Now we want to connect the model (\ref{model-rough}) with the mentioned
FMR experiments.
The probed quantity of an FMR experiment is the so called ``resonance
frequency'' that is identical to the spin wave mode with infinite wavelength
$\omega_0$.\cite{LiB03a} This mode is also called uniform ($q=0$) mode. In a coupled bilayer system there are two such excitations,
termed optical and acoustical mode ($\omega_{0o},\omega_{0a}$). This can
easily be understood if a semiclassical spin model is considered. The
uniform mode of a monolayer corresponds to a spin wave where all spins
rotate in phase. The acoustical mode of the coupled bilayer system corresponds to the state where the spins in both layers rotate in
phase. If the optical mode is excited, the spins in the upper layer
rotate in anti-phase compared to the spins of the lower layer. Depending
on the coupling sign, the optical mode can be found at higher
(ferromagnetic interlayer coupling) or at lower frequencies
(antiferromagnetic coupling). Note that
the FMR technique has equal access to ferromagnetic and
antiferromagnetic coupling. In this picture the distance between the
modes $\Delta \omega_0$
is a measure for the interlayer coupling strength if equivalent magnetic
layers are assumed.\\
In practice, during an FMR experiment the probe frequency $\nu_{hf}$ is
fixed in the microwave region, while the magnitude of the external field is varied until the
resonance frequency equals the probe frequency. 
\begin{eqnarray}
2\pi \nu_{hf}\stackrel{!}{=}\omega_{0(a,o)}(\vec{B}_0)
\label{Bres-def}
\end{eqnarray}
 This relation is fulfilled at special magnitudes of the external field
 called ``resonance fields'' ($B_{0r(a,o)}$). The values of the resonance
 fields as functions of the direction of the applied external fields
 ($B_{0r(a,o)}(\theta_{B_0})$) 
is the most important experimental output.\cite{Linienbreiten} For an
appropriately chosen probe frequency there are also two resonance
fields in a coupled bilayer system (\ref{model-rough}) belonging to the optical and acoustical mode. Again,
the distance between the resonance fields ($\Delta B_{0r}$) is
influenced by the interlayer coupling strength $J_I$ and the
anisotropies in both layers.   
 An FMR experiment can be analyzed using the model
(\ref{model-rough}) by calculating the spin wave modes
$\omega_{0(a,b)}$ and the resonance fields $B_{0r(a,o)}$ and fitting the model
parameters until the calculated values correspond to the measured
ones. It is possible to obtain the interlayer coupling as well as the
anisotropy strengths from the angular dependence of the resonance fields. 
\\  
 Let us now consider the temperature dependence of the resonance frequencies
 and fields.
There are two principal sources of temperature dependence:
\begin{enumerate}
\item
If the parameters of the model (\ref{model-rough}) are fitted to real
systems as proposed above, they may be temperature dependent itself. This applies especially to the interlayer coupling parameter $J_I$ which
is not a fundamental microscopic coupling constant but rather given in
Eq. (\ref{Ji}) and which may be significantly influenced by
temperature.\cite{Bru95, DKB99} As seen in Eq. (\ref{Ji}) the
temperature dependence of $J_I(T)$ describes
the temperature dependence of the spacer energies.
\item
Additionally, even if the model parameters are fixed, there is an intrinsic temperature dependence of
$\omega_{0(a,o)}$ and $\Delta B_{0(a,o)}$ in the model
(\ref{model-rough}) due to interactions of spin
waves excited within the magnetic layers. 
\end{enumerate}
It is obvious that the temperature dependence of the model parameter
$J_I(T)$ can be identified to the ``spacer effect'' (i) while the intrinsic
temperature dependence due to spin wave interactions is equivalent to the ``magnetic layer
effect'' (ii).\\
In this situation the microscopic model (\ref{model-rough}) can be used
to study the mechanisms (i) and (ii) separately. On the one hand one
can fit the model parameters to experimental FMR data which gives the
function $J_I(T)$. The latter exclusively describes the temperature
dependence of the IEC due to the ``spacer effect'' (i).\\
On the other hand the model parameters and especially $J_I$ can be
fixed. Then one can investigate the temperature dependence of the resonance frequencies due to spin wave
interactions. Thus the temperature dependence of the
FMR signal caused by the magnetic layer effect (ii) alone can be analyzed.\\
The motivation of our paper is twofold. First we want to show that an
evaluation of FMR data is in principle possible by use of the
microscopic model (\ref{model-rough}). However, our paper is mainly
dedicated to the investigation of the second point, i.e. of the
temperature dependence of the FMR signal caused by the interaction of
spin waves.  
\\
The paper is organized in the following way:\\
Using our microscopic model we will discuss two quantities: the resonance frequencies
$\omega_{0(a,o)}$ and the resonance fields $B_{0r(a,o)}$. We will consider a monolayer
(simulating a single film) as well as a system of two coupled layers
(which simulates prototype IEC-trilayer systems). We
compare the results of the classical model with those of the
microscopic model at zero temperature before we discuss the intrinsic
temperature dependence of the resonance frequencies and fields in
detail.
With this procedure we want to answer the following central questions:
\begin{itemize}
\item Is there a relevant intrinsic temperature dependence of the
  uniform spin wave modes in the
  model (\ref{model-rough}) and how is it characterized?
\item What are the differences between the results of our microscopic approach
  and of the macroscopic classical model for zero temperature, in
  other words, can the microscopic model give new insight in FMR
  experiments at a fixed temperature, too?
\end{itemize}
It turns out that the latter is indeed the case. For a nickel films
grown on a Cu(001) surface the absence of a resonance field in the hard
direction was reported.\cite{Lin02} This observation follows directly
for larger out-of-plane lattice anisotropies in the microscopic model,
while it can be hardly explained within the classical
model. Furthermore we will find a significant temperature dependence of the
uniform modes due to spin wave interactions. A description of this temperature dependence
in terms of effective model parameters ($\tilde{K}_2(T)$ and $\tilde{J}_I$(T)) is possible approximately only and justified only for lower temperatures $T<T_c$.      
 Before turning to the presentation
of our results we will shortly discuss the approximations used to solve
the model (\ref{model-rough}). The intended use poses clear requirements
to the theory. Its numerical effort must be moderate to keep the fitting procedure
feasible, while it has to yield accurate results for the spin waves for
arbritrary directions and magnitudes of the external field. Such a
theory is proposed and evaluated in Ref.~\onlinecite{SKN04} and applied to our
model (\ref{model-rough}) in the next section. As already mentioned this
contribution uses results and insights from earlier works, namely from
Ref. \onlinecite{ScN04} and Ref. \onlinecite{SKN04}. The former
(Ref. \onlinecite{ScN04}) discusses and compares the different mechanisms that may
cause a temperature dependence of the interlayer exchange coupling on a very
general level therewith clearly formulating the problem we want to
consider in this paper. The
latter (Ref. \onlinecite{SKN04}) is mere technically proposing a theory to solve our model (\ref{model-rough}). 

\section{theory}    
In this section we want to motivate the theoretical approach used to
solve the extended Heisenberg model (\ref{model-rough}).
 Due to the tetragonal symmetry the directions in the film plane are
equivalent. This allows us to restrict our calculations to a plane
perpendicular to the film, the $xz$-plane. The
external field and the magnetization are confined to this plane in the
following:
\begin{eqnarray}
\vec{B}_0&=&(B_{0x},0,B_{0z})\nonumber\\
\langle \vec{S}
\rangle&=&(\langle S_x\rangle,0,\langle S_z\rangle).\nonumber
\end{eqnarray}
To proceed we split the Hamiltonian (\ref{model-rough}) into two parts. The first one,
\begin{eqnarray}
H_1&=&
\sum_{\langle ij\rangle \alpha}J_{\alpha}\vec{S}_{i\alpha}\vec{S}_{j\alpha}
-\sum_{\langle i\alpha\,j\beta\rangle}^{\alpha\neq\beta} J_I\vec{S}_{i\alpha}\vec{S}_{j\beta}\nonumber\\
&&-\sum_{i\alpha}g_J\mu_B\vec{B}_0\vec{S}_{i\alpha}
-\sum_{i\alpha} K_{2\alpha}S_{i\alpha z}^2,
\label{H1}
\end{eqnarray} 
contains all terms except for the dipolar coupling. In a previous
paper\cite{SKN04} we developed a theory for the Hamiltonian (\ref{H1}) for the
special case of a monolayer. It was shown by comparison with Quantum
Monte Carlo (QMC) results\cite{HFK02} that our approximation is in fact
very accurate for arbritrary directions of the external field and can
therefore be used here. Our proposal combines an RPA decoupling procedure\cite{BoT59}
applied to the nonlocal Heisenberg exchange terms and a generalized
Anderson Callen approximation\cite{Cal63, AnC64} for the local
anisotropy term. Details are given in Ref.~\onlinecite{SKN04}. The theory is based on
a transformation into a local coordinate system $\Sigma\rightarrow
\Sigma^\prime$ where the new $z^\prime$-axis is parallel to the magnetization.
It is easy to
generalize this theory to a multilayer system if the RPA decoupling is
also used for the interlayer coupling
term. A remarkable result of this theory is that the
anisotropy and the interlayer coupling can be replaced by internal
fields $\vec{B}_{1\alpha}$ in the single
magnon Green function matrix of the transformed system:
\begin{eqnarray}
G_{\alpha\beta \vec{q}}^\prime(E)=\langle\langle
S_{\alpha\vec{q}}^{+\prime};S_{\beta -\vec{q}}^{-\prime}\rangle\rangle.
\label{GF-def}
\end{eqnarray}
The effective field is a sum of the external field, a term due to the
anisotropy and a further term resulting from the interlayer coupling:
\begin{eqnarray}
\vec{B}_{1\alpha}&=&g_{J\alpha}\mu_B
\vec{B}_{0}+\vec{B}_{K_2\,\alpha}+\vec{B}_{J_I\,\alpha}.
\label{eff-field1}
\end{eqnarray}
The interlayer field is given by
\begin{eqnarray}
\vec{B}_{J_I\,\alpha}=2J_Ip^\prime \langle \vec{S}_\gamma\rangle,
\label{JI-field}
\end{eqnarray}
where $p^\prime$ denotes the number of nearest neighbors in the adjacent
layer $\gamma$. For the anisotropy field one finds
\begin{eqnarray}
{B}_{K_2\,\alpha\,x}&=&-K_{2\alpha}\langle S_{\alpha x}\rangle \sin^2\theta_\alpha
C_1^\prime\nonumber\\
{B}_{K_2\,\alpha\,z}&=&+2K_{2\alpha}\langle S_{\alpha z}\rangle \left(1-\frac{1}{2}\sin^2\theta_\alpha\right)C_1^\prime
\nonumber\\
\mbox{with}\nonumber\\
C_1^\prime&=&1-\frac{1}{2S^2}\left(S(S+1)-\langle
    S_{\alpha z^\prime}^2\rangle\right),
\label{K2-field}
\end{eqnarray}
where $\theta_\alpha$ is the angle between the $z$-axis and the
magnetization in layer $\alpha$. Note that the anisotropy field
$\vec{B}_{K_2}$ is not an analytic function of the magnetization
$\langle S_{z^\prime} \rangle$ only, but of $\langle
S_{z^\prime}^2\rangle$-terms as well.
The single-magnon Green function
(\ref{GF-def}) is now a functional of the effective fields:
\begin{eqnarray}
\vec{G}_\vec{q}^\prime(E)=
\left(\begin{array}{cc}
 \langle 2S_{az^\prime}\rangle & 0\\
 0&\langle 2S_{bz^\prime}\rangle
\end{array}\right)\cdot \left(E \vec{I}-\vec{M_q}\right)^{-1},\nonumber
\end{eqnarray}
via  the matrix $\vec{M_q}$:
\begin{eqnarray}
{M}_\vec{q}^{\alpha\alpha}&=& \left(2J_\alpha\langle S_{\alpha z^\prime}\rangle (p-\gamma_\vec{q\|})-B_{1\alpha}\right)\nonumber\\
M_\vec{q}^{\alpha\beta}&=& \left(2J_I\langle S_{\beta z^\prime}\rangle
  \gamma_{\vec{q}\perp} \cos(\theta_a-\theta_b)\right)\nonumber\\
M_\vec{q}^{\beta\alpha}&=&(M_\vec{q}^{\alpha\beta})^\star.
\label{GF}
\end{eqnarray}
Generally, the magnetization angles in both layers, $\theta_a$ and
$\theta_b$, may be different. Note that the coordinate system is rotated
by $\theta_a$ in layer $A$ and by $\theta_b$ in layer $B$. Thus the
$z^\prime$-axes in both layers are not parallel to each other. The respective
angles are determined by the effective fields
\begin{eqnarray}
\tan\theta_{\alpha}=\frac{B_{\alpha x}}{B_{\alpha z}}
\label{angle}
\end{eqnarray}
Eq. (\ref{GF}) is valid as long as the deviations between the
magnetization angles of different layers are small.
\\
The coordination number within a layer is denoted by $p$ and the structural factors $\gamma_{\vec{q}\|}$ and $\gamma_{\vec{q}\perp}$
are a consequence of the two dimensional Fourier transformation and depend on the
lattice structure. For the fcc(100) geometry considered here
they are given by
\begin{eqnarray}
\gamma_{\vec{q}\|}&=&4\cos\frac{1}{2}aq_x\cos\frac{1}{2}aq_y\nonumber\\
\gamma_{\vec{q}\perp}&=&e^{-i\frac{1}{2}aq_x}\left(2\cos\frac{1}{2}aq_y+e^{-i\frac{1}{2}aq_x}\right),\nonumber 
\end{eqnarray}
where $a$ is the lattice constant.
\\
From the single-magnon Green function (\ref{GF}) one can easily obtain the expectation
values $\langle S_{z^\prime}\rangle$ and $\langle S_{z^\prime}^2\rangle$
using the standard text book procedure \cite{TaH62} which can be
generalized to film geometries straightforwardly.
\cite{ScN99} Additionally the spin wave frequencies
$h\nu_\vec{q}=E_\vec{q}$ are directly given by the eigenvalues
$E_\vec{q}$ of the matrix $\vec{M_q}$. Eqs. (\ref{eff-field1}) -
(\ref{angle}) establish a self consistent set of equations and thus the
problem is solved as far as the Hamiltonian (\ref{H1}) is concerned.\\ 
Now the dipolar term 
\begin{equation}
H_2=\sum_{ij\alpha}g_{0\alpha}\left(\frac{1}{r_{ij}^3}\vec{S}_{i\alpha}\vec{S}_{j\alpha}-\frac{3}{r_{ij}^5}
(\vec{S}_{i\alpha}\vec{r}_{ij})(\vec{S}_{j\alpha}\vec{r}_{ij})\right)
\label{H2}
\end{equation}
has to be treated. 
 The interaction is nonlocal and
very similar to the Heisenberg exchange terms. Thus an RPA decoupling
would be certainly a good choice. However it was shown in
Ref.~\onlinecite{FJK00a} that the more feasible mean field decoupling also yields
acceptable results. A
further term has to be added to the effective field (\ref{eff-field1})
due to the dipolar coupling:
\begin{equation}
  \vec{B}_{\alpha}=\vec{B}_{1\,\alpha}+\vec{B}_{{\rm dip}\,\alpha},\nonumber
\label{eff-field2}
\end{equation}
where the dipolar field is given by\cite{FJK00a}
\begin{eqnarray}
B_{{\rm dip}\,\alpha\,x}&=&\quad\; g_{0\alpha}\langle S_{x\alpha}\rangle \sum_{j}\frac{1}{r_{0j}^{3}}\nonumber\\ 
B_{{\rm dip}\,\alpha\,z}&=&-2 g_{0\alpha}\langle S_{z\alpha}\rangle \sum_{j}\frac{1}{r_{0j}^{3}}.
\label{dipol-field}
\end{eqnarray}
The sum runs over all lattice sites of a layer.
There
are no further changes in the relations (\ref{eff-field1})-
(\ref{angle}) due to the dipolar coupling.
Thus the effective field $\vec{B}_\alpha$ defined in
Eqs. (\ref{eff-field2}) and (\ref{dipol-field}) simply
has to replace the field $\vec{B}_{1\alpha}$ from Eq. (\ref{eff-field1}) if one
wants to account for dipolar coupling.\\
As already mentioned before, this theory is capable of calculating the
whole spin wave excitation spectrum and especially the uniform modes $\omega_{0o}$ and $\omega_{0a}$. Therewith the
resonance fields $B_{0r(a,o)}$ can be obtained for any microwave probe frequency
$\nu_{hf}$. This holds for arbritrary external field directions and for
any temperature. Thus one can calculate the central functions $B_{0r(a,o)}(\theta_{B_0})$
and $B_{0r(a,o)}(T)$ as well as the dependence of the resonance frequency on
the external field $\omega_{0(a,o)}(B)$. The latter is often referred to as
"dispersion relation" in connection with FMR experiments and gives 
the excitation gap $E_{\rm gap}=\hbar\cdot {\rm min}(\omega_{(0a)},\omega_{(0o)})$.
The results will be
presented and discussed in the next section.   

\section{The single film}
\begin{figure}
\epsfig{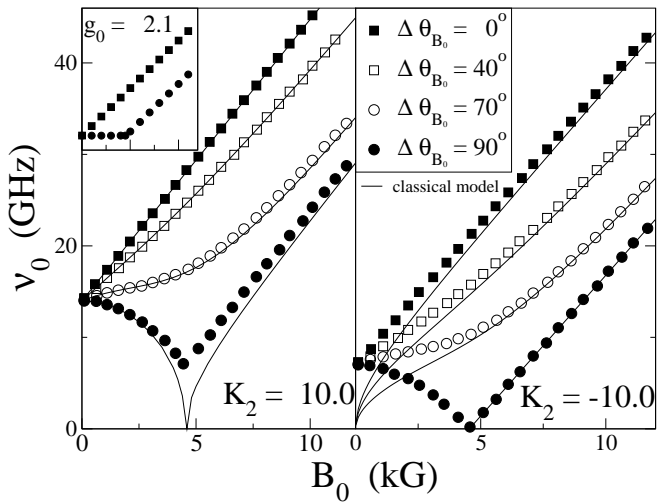}
\caption{The resonance frequency as a
  function of the external field at $T=0$ for different angles between
  the easy direction and the external
  field $\Delta
  \theta_{B_0}$ (symbols). Left panel: positive lattice anisotropy
  $K_2=10\, \mu_B {\rm kG}$, right panel: negative lattice anisotropy
  $K_2=-10\, \mu_B {\rm kG}$, inset: dipolar
  coupling $g_0=2.1\, \mu_B {\rm kG}$. Here and in the following pictures the
  spin quantum number is set to unity (S=1). The results
  of the classical model are also shown (solid lines).} 
\label{Fig1}
\end{figure}

\begin{figure}
\epsfig{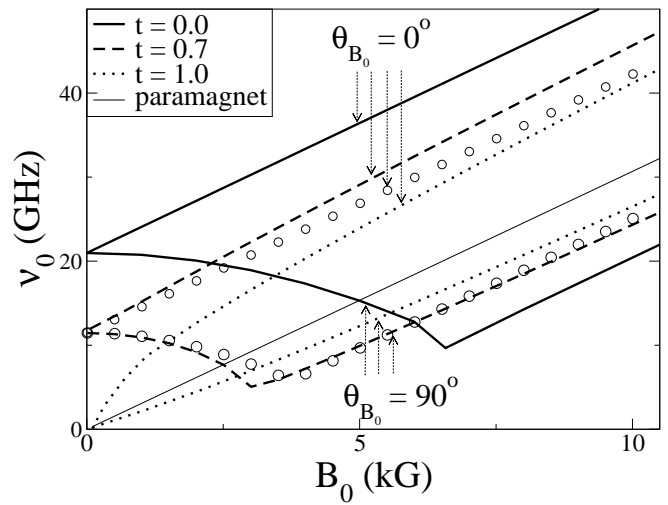}
\caption{The dispersion relations $\nu_0(B_0)$ at four different reduced
  temperatures $t=T/T_c$. The external field is applied in the easy
  (upper curves) and in the hard direction (lower curves). The lattice
  anisotropy is $K_2=15\, \mu_B {\rm kG}$ and the dipolar coupling is zero. The thin
  solid line gives the resonance frequencies for
  paramagnetism. Circles: $t=0,\, K_2=8.2\, \mu_B {\rm kG}$} 
\label{Fig2}
\end{figure}

In the following sections the anisotropy parameters and the dipolar coupling will
be given in units of $[\mu_B{\rm kG}]$ with 
$1\, \mu_B{\rm kG}=5.79\, {\rm \mu eV}$. As can be seen in the Hamiltonian~(\ref{model-rough}) the parameters $J_I, J_\alpha, K_2$ and $g_0$ denote
energies per lattice site, i.e. energy densities rather than energies.  
The external fields are given in $[{\rm kG}]$ while the resonance frequencies
$\nu_0=\omega_0/2\pi$ are given in $[{\rm GHz}]$. The spin quantum number is
set to unity (S=1) and the Lande factor is fixed to the bulk nickel value $g_{J\alpha}=2.21$. \\
Though we are mainly interested in the coupled bilayer system one has to
understand the single film system before a meaningful interpretation of more complex
systems is possible. Moreover, though the experimental output is the resonance field
it is very instructive to study the ``dispersion relation''. The dispersion relations for zero
temperature are shown in Fig. \ref{Fig1}. The spin wave frequency
$\nu_0$ is displayed as a function of the external field
$B_0$ for positive and
negative lattice anisotropies $K_2=\pm 10\,\mu_B{\rm kG}$ (left and right
panel, $g_0=0$) as well as for
dipolar coupling
$g_0=2.1\,\mu_B{\rm kG}$ (inset, $K_2=0$).
The angle
$\Delta \theta_{B_0}$ denotes the angle between the external field and
the easy direction. The latter is perpendicular to the film for a positive lattice anisotropy
and parallel to the film plane for negative
lattice anisotropies or for dipolar coupling. For comparison the results of the
classical model as e.g. given in Ref.~\onlinecite{LiB03} are also shown
(thin solid lines). The
anisotropy parameter $\tilde{K}_2$ used for the classical calculation
is scaled as proposed in Ref.~\onlinecite{FJK00} ($\tilde{K_2}=S(S-1/2)K_2$) and accounts for both,
lattice and dipolar anisotropies.
Qualitatively the results are similar for all cases. If the
external field is applied parallel to the easy direction ($\Delta
\theta_{B_0}=0^\circ$) the function $\nu_0(B_0)$
increases linearly. The
intersection with the ordinate gives the minimal excitation gap due to the
anisotropy. This gap is larger for positive than for negative
lattice anisotropies and rises with the norm of $K_2$ as can be seen by
a comparison of the left panel of Fig. \ref{Fig1} ($K_2=10\, \mu_B {\rm kG}$)
and the solid line in Fig. \ref{Fig2} ($K_2=15\,\mu_B {\rm kG}$).
 There are more features in the dispersion
relation if the external field is applied in the hard
direction ($\Delta
\theta_{B_0}=90^\circ$). The magnetization is now rotated out of its easy direction
and finally aligned parallel to the external field if the latter exceeds
a certain value called reorientation field.\cite{SKN04, JeB98} If the
external field is enhanced further the
magnetization does not change its orientation any more and the spin
wave frequency $\nu_0$ is again a linearly increasing function of the external
field. However, before the reorientation field is reached, $\nu_0(B_0)$
is constant (for dipolar coupling) or decreases (for
lattice anisotropies) as a function of the external field. There is an
interesting feature for negative $K_2$: at the reorientation field the
resonance frequency is zero, i.e. the excitation gap is closed in spite
of the presence of an external field and lattice anisotropies. This is
not found for positive values of $K_2$ or for dipolar coupling, where
the gap stays finite at the reorientation field (left panel).
 In such cases the probe frequency $\nu_{hf}$ may be
lower than the resonance frequency $\nu_0$ for all angles
$\Delta\theta_{B_0}$ and no resonance field is detected at all. This feature
is not reproduced by the classical model, but experimental evidence for
such a scenario was indeed reported for nickel
films with enhanced uniaxial anisotropy.\cite{Lin02}\\
There are more differences between the classical and quantum mechanical
theory for negative anisotropies and dipolar coupling, namely for field
values that are smaller than the reorientation field. Consider, for example, an external field applied
perpendicularly to the easy direction ($\theta_{B_0}=90^\circ$,
dots). Here the resonance frequency and therewith the excitation gap is
zero for all field values below the reorientation field in the classical calculation. The same is expected for the
quantum mechanical calculation, since in this region the magnetization
has a finite component parallel to the film plane in the ground state and
thus the ground state is
degenerated with respect to the azimutal angle $\phi$ of the
magnetization.
 Therefore a Goldstone mode is expected, i.e. the
spin-wave energy for $q=0$ should be zero. This feature is not reproduced by
our approximation, since it restricts the problem to a plane
perpendicular to the film plane.     
However, despite these differences, the agreement between the classical and
the microscopic theory is rather good for a large region of the
parameter space at $T=0$. 
\\ Let us now consider the
temperature dependence of the dispersion relation. In the following
discussion the external field will be applied either along the easy
direction or along the hard direction. This configurations will be
referred to shortly as ``easy direction'' or ``hard direction'' case.
\begin{figure}
\epsfig{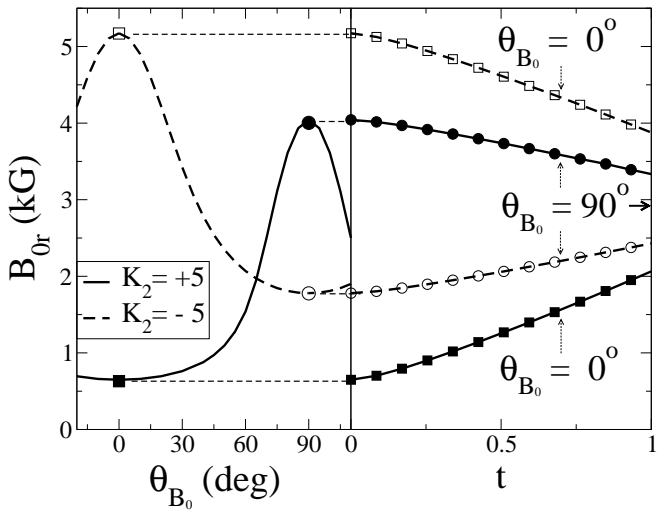}
\caption{The resonance fields for positive (solid lines, filled symbols) as well as for
  negative (dashed lines, open symbols) lattice anisotropies as a function of the
  polar angle of the external field (left panel, $T=0$) and as a function of
  reduced temperature $t=T/T_c$ (right panel,
  $\theta_{B_0}=0^\circ$ - squares, $\theta_{B_0}=90^\circ$ - circles). The position of the
  resonance field of a paramagnet is indicated by the arrow in the
  right panel.} 
\label{Fig3}
\end{figure}
In Fig. \ref{Fig2} we
display the dispersion relation for positive $K_2$ at three different
reduced temperatures $t=T/T_c$ ranging from $t=0.0$ to $t=1.0$ for the
easy as well as for the hard direction case.  Additionally the result
for a reduced anisotropy $K_{2}^\prime=0.547K_2$ at zero temperature
$t=0$ is shown (circles). The thin solid line denotes the resonance as found for a
paramagnet ($J_\alpha=0$). The dispersion relation shows a significant temperature dependence:
With increasing $t$ the curves move towards the paramagnetic line. The
{\it magnitude} of this shift in a certain temperature interval is determined by the
distance between the spin wave frequency at $t=0$ and the
paramagnetic mode. The resonance frequencies for the easy
direction case are more sensitive to temperature than those for the hard
direction case, since the former are further away from the paramagnetic line.
 However, the paramagnetic solution is not
reached at $T_c$ since a finite magnetization is induced by the
external field. Further one notices that the shape of the curves at
$t=0$ and at $t=0.7$ are qualitatively the same. Temperature 
reduces the resonance frequency for zero external field as well as the
reorientation field. Similar effects are found for fixed temperature
 but reduced anisotropy constant $K_2$ as can be
seen by comparing the results for $K_2=15\,\, \mu_B{\rm kG}, t=0.7$ (dashed
line) and $K_2=8.2\,\mu_B{\rm kG},\, t=0$ (circles). However, there are still
differences between both cases and hence a reduced anisotropy parameter
has clearly not the same effects on the dispersion relation as increasing
temperature. 
The curves for temperatures in the vicinity or
above the Curie temperature and for $t=0$ differ even qualitatively. 
\begin{figure}
\epsfig{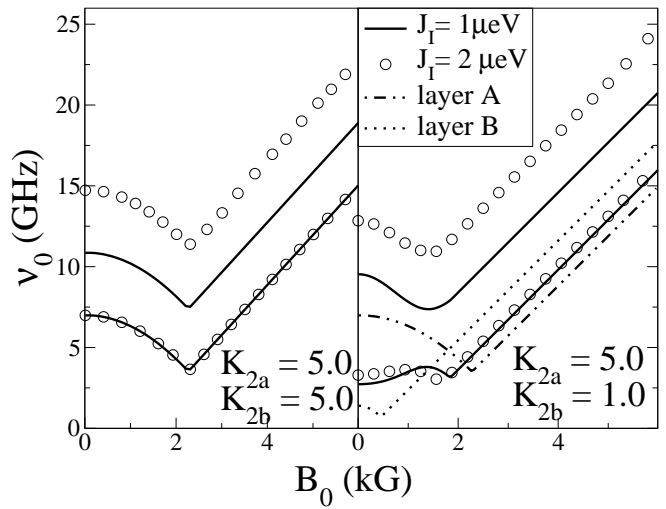}
\caption{The resonance frequencies as a function of the external field
  applied in the hard direction for a coupled bilayer system. Left:
  symmetric film system, $K_{2a}=K_{2b}=5\,\mu_B{\rm kG}$,
 right: Asymmetric film system $K_{2a}=5K_{2b}=5\,\mu_B{\rm kG}$. For both
 panels the interlayer coupling strengths are $J_I=1\,{\rm \mu
 eV}$ (solid lines) and $J_I=2\, {\rm \mu eV}$ (circles). Additionally the results
 for single films, i.e. $J_I=0$ are shown in the right panel(dash-dotted line - layer A and dotted
 line - layer B).}
\label{Fig4}
\end{figure}
Similar results are found for negative lattice anisotropies and for dipolar coupling.
These findings bear important consequences for the usual interpretation of FMR
data in terms of a classical model where the effect of temperature is
taken into account via effective anisotropy
parameters $\tilde{K}_2(T)$. Obviously, this ansatz is an acceptable approximation
only for lower temperatures and not at all justified for temperatures in
the vicinity or above $T_c$.
\\
At the end of this section we want to discuss resonance fields (Fig.
\ref{Fig3}). The angular as well as the temperature dependence is shown
for positive (solid line, filled symbols) as well as for
negative (dashed line, open symbols) lattice
anisotropies. A probe frequency of $9\,{\rm GHz}$ is chosen. It is clearly seen that the resonance field has a maximum
at the hard direction and a minimum at the easy direction. The
resonance fields at these extremal points are larger for negative
 than for positive anisotropies.\\
If the temperature is enhanced, the resonance fields shift towards the
paramagnetic limit $B_{0r}({\rm param}) = 2.9\,{\rm kG}$, which is however not
reached before $t\approx 2.5$. Again the
{magnitude} of these
shifts in a certain temperature interval scale with the distance between the resonance field at zero
temperature and the
resonance field found for a paramagnet. Thus for both signs of the
anisotropy the temperature dependence of the resonance field is more pronounced if the external field is
applied perpendicularly to the film plane rather than  parallely.
The resonance fields increase
with temperature in the easy direction 
and decrease with temperature in the hard direction. All characteristic
features follow from the respective
dispersion relations in Fig. \ref{Fig1} and Fig. \ref{Fig2}. 
Keeping in
mind the behavior of the single film we now want to analyze the coupled
bilayer system to understand the intrinsic temperature dependence of the
interlayer exchange coupling.
Concerning the coupled system we will consider two cases: a symmetric
film system where the upper and the lower layers are equivalent and an
asymmetric film system, i.e. a sytem with different anisotropy strengths in the upper and
the lower layer.
\begin{figure}
\epsfig{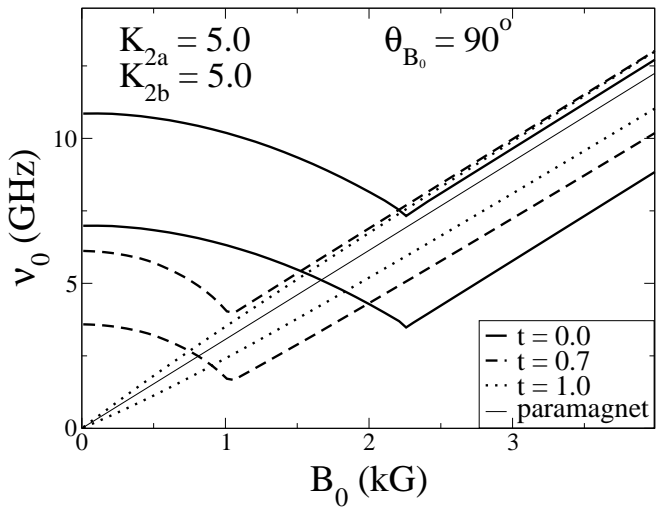}
\caption{The dispersion relations for a symmetric film system
  ($K_{2(a,b)}=5\,\mu_B{\rm kG},\, J_I=1\,{\rm \mu eV}$) for different reduced temperatures
  $t=T/T_c$. The external field is applied in the hard direction. The paramagnetic
  solution is also shown (thin solid line).}
\label{Fig5}
\end{figure}
\section{The interlayer coupling}
We start our discussion considering dispersion relations.
Fig. \ref{Fig4} shows the dispersion relation of two equivalent layers
(left panel, $K_{2a}=K_{2b}=5\,\mu_B{\rm kG}$)
as well as of an asymmetric film system (right panel, $K_{2a}=5\,\mu_B{\rm kG},\,K_{2b}=1\,\mu_B{\rm kG}$). The interlayer
couplings are $1\, {\rm \mu eV}$ (solid lines) and $2\, {\rm \mu
eV}$ (circles), the temperature is zero. For the asymmetric system the
results for decoupled films ($J_I=0$) is also shown. 
 The easy direction is
perpendicular to the film and the external field is applied in the hard
direction. For the coupled film two modes are found, the acoustical
(lower) and
the optical one (upper). In the case of symmetric parameters the curves are parallel and the distance between the
modes $\Delta \nu_0$ scales with the interlayer coupling strength
$J_I$. The position of the acoustical mode is not influenced by the
interlayer coupling strength. The same applies for the easy direction case.
The situation is somewhat more
complex if the layers are not equivalent as shown in the right panel.
For $B_0=0$
the modes have lower frequencies compared to the symetric system. This
is caused by the reduced anisotropy strenght in layer $b$. Furthermore the modes are not parallel any more and the
distance between the modes depends on both: the film asymmetry and the
interlayer coupling. Other than for symmetric films the acoustical mode is now slightly
influenced by the interlayer coupling strength $J_I$.
 It is located between the modes of the upper and the lower layer of the
 decoupled system. 
\begin{figure}
\epsfig{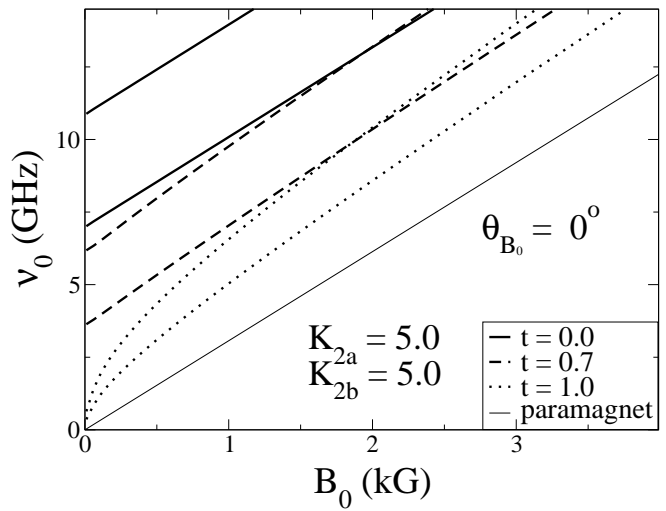}
\caption{The same as Fig. \ref{Fig5}, but for an external field applied in
  the easy direction ($\theta_{B_0}=0^\circ$).}
\label{Fig6}
\end{figure}
 \\ Next we want to
analyze the temperature dependence of the resonance modes at fixed
anisotropy and interlayer coupling parameters. Note that
this is not possible within the classical theory,\cite{SmB55, Kit47}
where different anisotropy parameters and interlayer coupling constants
have to be assumed for every temperature. 
Fig. \ref{Fig5} and Fig. \ref{Fig6} show the dispersion relations for equivalent layers at
three different reduced temperatures: $t=0.0$, $t=0.7$ and $t=1.0$ for
the easy and the hard direction case as well as the dispersion relation
of a paramagnet (thin solid lines). 
Similar trends as for the single layer are found for the temperature
dependence of the resonance frequencies.   
All modes move towards the paramagnetic solution with increasing
temperature and the magnitudes of the shifts scale with the
distance to the paramagnetic mode. For the hard direction case
(Fig. \ref{Fig5}) the optical mode is influenced only slightly in the
linear regime, since the distance to the paramagnetic mode is very
small. Because in this regime this distance is larger for the acoustical
mode the latter is
stronger influenced by temperature.
For zero external field the mode frequencies of the optical and
acoustical mode get smaller with increasing temperature and reach zero
at $T_c$. 
The distance between the optical and acoustical mode is clearly reduced
by increasing temperature but stays finite for finite fields at the
Curie temperature.
For the easy direction case (Fig.~\ref{Fig6}) the $t=0$ resonance frequencies
of both modes are clearly larger than the paramagnetic one. This causes a
significant decrease of the resonance frequencies with
increasing temperature.
The distance between the modes $\Delta \nu_0$ is reduced with temperature. This reduction
is comparable in the easy and in the hard direction
case.
\begin{figure}
\epsfig{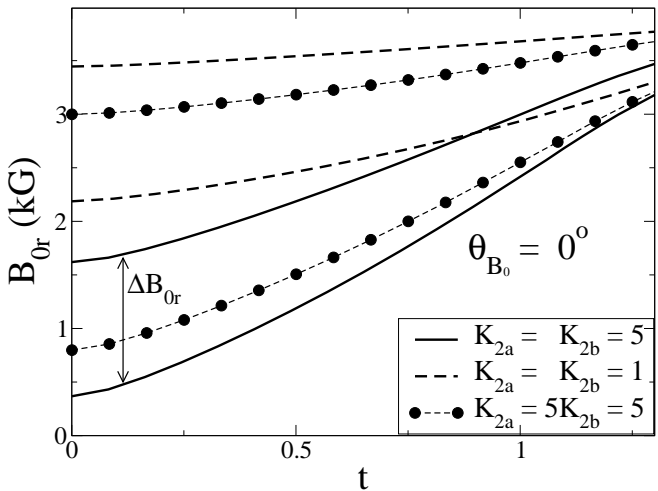}
\caption{The resonance fields as a function of temperature for the easy
  direction. The results of two symmetric film systems
  ($K_{2a}=K_{2b}=5\,\mu_B{\rm kG}$ -
  solid lines, $K_{2a}=K_{2b}=1\,\mu_B{\rm kG}$ -
  dashed lines) as
  well as of an asymmetric film system ($K_{2a}=5K_{2b}=5\,\mu_B{\rm kG}$ -
  dashed lines with circles) are shown. Further parameters: $J_I=1 {\rm \mu
  eV},\, \nu_0=12 {\rm GHz}$.}
\label{Fig7}
\end{figure}
\\
Let us now consider the temperature dependence of the resonance fields
for the coupled bilayer system. The probe frequency is chosen to be
$12\,{\rm GHz}$. The resonance field for a paramagnet is $ B_{\rm 0r}({\rm
  param})= 3.87\, {\rm kG}$.
Fig.~\ref{Fig7} (easy direction) and Fig.~\ref{Fig8} (hard direction) show the resonance fields as a function of temperature
for ferromagnetic interlayer coupling ($J_I=1{\rm \mu eV}$). Two symmetric
films with higher ($K_2=5\,\mu_B{\rm kG}$ - solid
lines) and smaller anisotropy strength ($K_2=1\,\mu_B{\rm kG}$ - dashed lines) as well as an asymmetric film
system (symbols) are considered. 
In the
easy direction case all resonance fields are smaller than the
paramagnetic solution and therefore the fields increase with temperature. For zero temperature
the resonance fields are smaller for larger anisotropy strengths. The
mode distance, however, is the same for both symmetric systems ($\Delta
B_{0r} \approx 1.25\, {\rm kG}$). It is clearly seen that the asymmetric system
is somehow a "mixture" between the considered symmetric
systems.
\begin{figure}
\epsfig{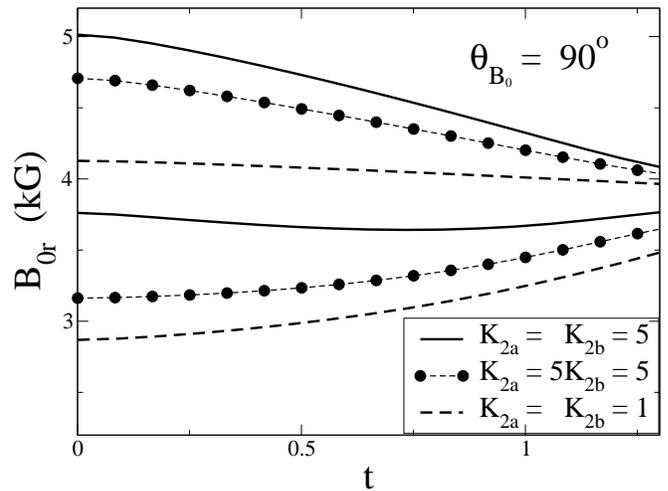}
\caption{The same as Fig. \ref{Fig7}, but for external fields applied in
the hard direction.}
\label{Fig8}
\end{figure}
 The acoustical as well as the optical resonance fields are located between
the respective fields of the symmetric systems. The resulting distance between the optical and the
acoustical field is larger than for the symmetric systems. This is in
agreement to the findings of Fig. \ref{Fig4}. Since the resonance fields
for higher anisotropies are further away from the paramagnetic solution
they react more sensitively to temperature than the resonance fields for the
low anisotropy system. The distance between the modes is reduced by
temperature for all considered systems.\\
The behavior in the hard direction case is similar. The acoustical
resonance fields are larger than the paramagnetic one and consequently
the fields are reduced by temperature. The resonance fields of the
optical mode increase with temperature for the low anisotropy and for
the asymmetric film system since they are smaller than the paramagnetic
field. If the resonance field is very close to the paramagnetic one the
resonance field may behave non monotonically as found for the optical
resonance of the 
high anisotropy system (lower solid line). As for the easy direction case the distance
between the optical and acoustical resonance field $\Delta B_{0r}$ is
reduced by temperature. 
The last point is of special interest for symmetric
systems. Recall that for zero temperature as well as within the classical
approach\cite{SmB55, Kit47} the distance between the
optical and acoustical field $\Delta B_{0r}$ is solely determined by the
interlayer coupling strength. 
This feature, however, is lost at finite
temperatures. As can be seen in Fig.~\ref{Fig9} the distance shrinks
faster for the high anisotropy system $K_2=5\,\mu_B{\rm kG}$ than for the
system with lower anisotropy $K_2=1\,\mu_B{\rm kG}$, irrespective of the same
interlayer coupling strength. Thus the field
distance $\Delta B_{0r}$ depends on both, on the interlayer coupling
strength as well as on the anisotropies. Note that for finite
temperatures this holds even for symmetric
systems, i.e. systems with equivalent magnetic films. This behavior is
not reproduced by the classical theory\cite{SmB55, Kit47}.
For the asymmetric system the
temperature dependence is comparable to those of  the high anisotropy
system. Furthermore the temperature dependence of the 
distance between the optical and acoustical field $\Delta B_{0r}$ 
is slightly enhanced for higher coupling strengths (open squares).
\\
Let us summarize our findings for the coupled bilayer system:
\begin{itemize}
\item There are two uniform modes belonging to the acoustical and
  optical spin wave branch for the coupled bilayer system. For equivalent
  layers and at $t=0$ the mode distance $\Delta \nu_0$ is a measure for the
  interlayer coupling strength.
\item Both modes and the respective resonance fields are shifted towards the paramagnetic solution with
  increasing temperature. The magnitude of these shifts is determined by
  the distance of the respective mode from the paramagnetic line. The
  distance between the optical and acoustical mode $\Delta \nu_0$ and the distance between the resonance
  fields $\Delta B_{0r}$ are reduced by temperature.
\item The magnitudes of these reductions are found to be larger for
  systems with a larger anisotropy strength.
\end{itemize}
The temperature dependence of the resonance frequencies and fields is
reflected in the temperature dependence of the effective interlayer
coupling $\tilde{J}_I(T)$ if the latter is obtained from experiment. It is clear that the
effective interlayer coupling $\tilde{J}_I(T)$
decreases if the temperature is enhanced. The magnitude of this decrease, however, may
depend on the anisotropies of the film and is larger for larger anisotropies.
 Note that this
behavior is exclusively due to the contribution of spin wave interactions to
the temperature dependence of the uniform
mode.

\begin{figure}
\epsfig{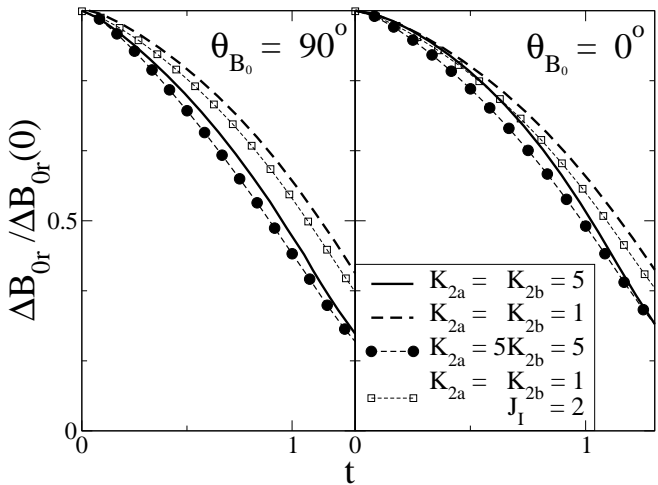}
\caption{The distance between the optical and acoustical resonance field
$\Delta B_{0r}$ as a function of the reduced temperature $t=T/T_c$ 
normalized to its $t=0$ value. For $J_I=1\, {\rm \mu eV}$ results are shown for a symmetric system
with a lower anisotropy (dashed line), for a symmetric system with a
higher anisotropy (solid line) and an asymmetric film system with a high
anisotropy layer and a low anisotropy layer (filled dots). For $J_I=2\,
{\rm \mu eV}$ results are shown for the symmetric low anisotropy system (open squares).}
\label{Fig9}
\end{figure} 
\section{Conclusions and Outlook}
Commonly there are two methods used to evaluat FMR experiments, namely
a method based on an expansion of the free energy\cite{SmB55} or a
method based on the equation of motion for the classical magnetization
vector.\cite{Kit47} Both techniques use a classical continuum model. In
this paper we
propose an alternative method based on an extended microscopic
Heisenberg model. The theoretical treatment of this model was shown to
give excellent results for arbritrary directions of the external
field.\cite{SKN04} 
This new approach allows for an explicit consideration
of temperature as far as magnetic excitations are concerned. We used
the theory to make general predictions about the temperature dependence of the FMR frequencies and fields caused by
magnetic excitations. Additionally the results of the new method were
compared to results of the classical model. A single film as well as an
interlayer coupled bilayer system were considered. We found the following
results:
\begin{itemize}
\item For zero temperature the results of the new method are very similar to those of the
  classical model for many parameter settings. There is however an
  important difference in the hard
  direction: The resonance frequency stays finite for all external
  fields for positive lattice anisotropies, while the classical model predicts a vanishing resonance
  frequency at the reorientation field. That is why the classical
  model can not explain the absence of a resonance field in the hard
  direction which was, however, observed experimentally for some
  systems.\cite{Lin02}  Using the microscopic analysis
  this observation turns out to be the natural consequence of the
  finite frequency minimum in the dispersion relation (see
  left panel of Fig. \ref{Fig1}). For large
  anisotropy parameters the probe frequency is lower than the minimal
  resonance frequency 
  and no resonance field can be detected any more.
\item For the coupled bilayer system two modes 
  are found. For $t=0$ and equivalent magnetic layers the
  distance between the modes is solely determined by the interlayer
  coupling strength. 
For a system with non-equivalent layers
  the mode distance depends on the interlayer coupling as well as on the
  asymmetry of the film system. 

\item The resonance frequencies and fields have a significant
  temperature dependence. The dispersion
  relation looks qualitatively similar for temperatures below $T_c$. The
  frequency for zero external field as well as the minimum of the dispersion
  relation are shifted to lower values with increasing $t$. This effects are similar, but not equivalent, to a reduction
  of the anisotropy parameter (see Fig. \ref{Fig2}). For temperatures in
  the vicinity or above $T_c$ the shape of the dispersion relation
  changes drastically compared to $t=0$.
\item With increasing temperature the resonance frequencies and fields
  are shifted towards the paramagnetic solution. This applies for the optical as well as
  for the acoustical branch and causes a decrease of the distance
  between the optical and acoustical mode $\Delta \nu_0$ as well as of
  the distance between the respective resonance fields $\Delta B_{0r}$. 
  However, the paramagnetic solution is not reached at $T_c$ since the
  external field induces a finite magnetization above $T_c$. 
\item For symmetric films the behavior of $\Delta B_{0r}$ is noteworthy. At $t=0$ it is
  solely determined by the interlayer coupling strength $J_I$. However, the anisotropy parameters do
  influence the temperature dependence of $\Delta B_{0r}$. The latter
  decreases faster with temperature for larger anisotropies. This
  qualitatively different behavior of the mode distance for zero and
  finite temperatures can not be described by the classical approach.   
\end{itemize}
These are the trends concerning the temperature dependence of the FMR
signals due to the interaction of spin waves.\\
Furthermore we expect valuable insights, if our model is used to
analyze temperature dependent FMR measurements.
Concerning the temperature dependence of the
interlayer exchange coupling this method can be used to separate the
different mechanisms which are known to be present in such systems,
namely the spacer effect due to a smearing out of the Fermi surface
within the spacer and the magnetic effect caused by the excitation and
interaction of spin waves. The relevance of both effects is intensively
discussed in literature but a separation and thus a direct comparison of
both effects was not possible up to now. Such calculations
are intended for the future.
Additionally the procedure
can be used to analyze temperature dependent FMR measurements of
anisotropy parameters.\cite{Far98} Again the temperature dependence due to spin
waves can be separated from other sources of temperature dependence
which opens the possibility to gain a deeper understanding of anisotropy
effects, too.

\section*{Acknowledgments}
This work is supported by the Deutsche Forschungsgemeinschaft within
the Sonderforschungsbereich 290. Productive discussions with
K. Baberschke and his group are gratefully acknowledged.


\begin{thebibliography}{10}
\bibitem{Linienbreiten}
of course also other quantities, as e.g. the linewidth are an
interesting experimental output. However, in this paper we will
exclusively consider resonance fields and frequencies.
\end{thebibliography}
\end{document}